\documentclass[preprint,5p,twocolumn]{elsarticle}   %final style
\biboptions{sort&compress}

\usepackage{lineno,hyperref}
\modulolinenumbers[4]

\usepackage{amssymb}
\usepackage{amsmath}                                    % this is for flalign to aling formulars
\usepackage{booktabs}                                    % Make nice tables
\usepackage{floatrow}
\usepackage{pst-all}                                         % ps tricks 

\newcommand{\KNN}{$\overline{K}NN$ }
\newcommand{\KNNo}{$\overline{K}NN$}
\newcommand{\pkl}{$pK^{+}\Lambda$ }
\newcommand{\pklo}{$pK^{+}\Lambda$}
\newcommand{\pl}{$p\Lambda$ }
\newcommand{\plo}{$p\Lambda$}
\newcommand{\KN}{$\overline{K}N$ }

\newcommand{\MeV}{MeV/c$^{2}$ }
\newcommand{\MeVo}{MeV/c$^{2}$}

\begin{document}
\begin{frontmatter}
\title{Partial Wave Analysis of the Reaction $p(3.5 GeV)+p \to pK^+\Lambda$ to Search for the "$ppK^-$" Bound State}
%
%\collaboration{HADES Collaboration}
%
\author[7]{G.~Agakishiev}
\author[9,10]{O.~Arnold}
\author[18]{D.~Belver}
\author[7]{A.~Belyaev}
\author[9,10]{J.C.~Berger-Chen}
\author[2]{A.~Blanco}
\author[10]{M.~B\"{o}hmer}
\author[16]{J.~L.~Boyard}
\author[18]{P.~Cabanelas}
\author[7]{S.~Chernenko}
\author[3]{A.~Dybczak}
\author[9,10]{E.~Epple\corref{mycorrespondingauthor}}
\author[9,10]{L.~Fabbietti\corref{mycorrespondingauthor}}
\author[7]{O.~Fateev}
\author[1]{P.~Finocchiaro}
\author[2,b]{P.~Fonte}
\author[10]{J.~Friese}
\author[8]{I.~Fr\"{o}hlich}
\author[5,c]{T.~Galatyuk}
\author[18]{J.~A.~Garz\'{o}n}
\author[10]{R.~Gernh\"{a}user}
\author[8]{K.~G\"{o}bel}
\author[13]{M.~Golubeva}
\author[5]{D.~Gonz\'{a}lez-D\'{\i}az}
\author[13]{F.~Guber}
\author[5,c]{M.~Gumberidze}
\author[4]{T.~Heinz}
\author[16]{T.~Hennino}
\author[4]{R.~Holzmann}
\author[7]{A.~Ierusalimov}
\author[12,e]{I.~Iori}
\author[13]{A.~Ivashkin}
\author[10]{M.~Jurkovic}
\author[6,d]{B.~K\"{a}mpfer}
\author[13]{T.~Karavicheva}
\author[4]{I.~Koenig}
\author[4]{W.~Koenig}
\author[4]{B.~W.~Kolb}
\author[5]{G.~Kornakov}
\author[6]{R.~Kotte}
\author[17]{A.~Kr\'{a}sa}
\author[17]{F.~Krizek}
\author[10]{R.~Kr\"{u}cken}
\author[3,16]{H.~Kuc}
\author[11]{W.~K\"{u}hn}
\author[17]{A.~Kugler}
\author[10]{T.~Kunz}
\author[13]{A.~Kurepin}
\author[7]{V.~Ladygin}
\author[9,10]{R.~Lalik}
\author[9,10]{K.~Lapidus}
\author[14]{A.~Lebedev}
\author[2]{L.~Lopes}
\author[8]{M.~Lorenz}
\author[10]{L.~Maier}
\author[2]{A.~Mangiarotti}
\author[8]{J.~Markert}
\author[11]{V.~Metag}
\author[8]{J.~Michel}
\author[8]{C.~M\"{u}ntz}
\author[9,10]{R.~M\"{u}nzer}
\author[6]{L.~Naumann}
\author[8]{Y.~C.~Pachmayer}
\author[3]{M.~Palka}
\author[15,f]{Y.~Parpottas}
\author[4]{V.~Pechenov}
\author[8]{O.~Pechenova}
\author[4]{J.~Pietraszko}
\author[3]{W.~Przygoda}
\author[16]{B.~Ramstein}
\author[13]{A.~Reshetin}
\author[8]{A.~Rustamov}
\author[13]{A.~Sadovsky}
\author[3]{P.~Salabura}
\author[a]{A.~Schmah}
\author[4]{E.~Schwab}
\author[10]{J.~Siebenson}
\author[17]{Yu.G.~Sobolev}
\author[g]{S.~Spataro}
\author[11]{B.~Spruck}
\author[8]{H.~Str\"{o}bele}
\author[8,4]{J.~Stroth}
\author[4]{C.~Sturm}
\author[8]{A.~Tarantola}
\author[8]{K.~Teilab}
\author[17]{P.~Tlusty}
\author[4]{M.~Traxler}
\author[15]{H.~Tsertos}
\author[7]{T.~Vasiliev}
\author[17]{V.~Wagner}
\author[10]{M.~Weber}
\author[6,d]{C.~Wendisch}
\author[6]{J.~W\"{u}stenfeld}
\author[4]{S.~Yurevich}
\author[7]{Y.~Zanevsky} 
\author[]{\\
(HADES Collaboration)}
\address[1]{Istituto Nazionale di Fisica Nucleare - Laboratori Nazionali del Sud, 95125~Catania, Italy}
\address[2]{LIP-Laborat\'{o}rio de Instrumenta\c{c}\~{a}o e F\'{\i}sica Experimental de Part\'{\i}culas , 3004-516~Coimbra, Portugal}
\address[3]{Smoluchowski Institute of Physics, Jagiellonian University of Cracow, 30-059~Krak\'{o}w, Poland}
\address[4]{GSI Helmholtzzentrum f\"{u}r Schwerionenforschung GmbH, 64291~Darmstadt, Germany}
\address[5]{Technische Universit\"{a}t Darmstadt, 64289~Darmstadt, Germany}
\address[6]{Institut f\"{u}r Strahlenphysik, Helmholtz-Zentrum Dresden-Rossendorf, 01314~Dresden, Germany}
\address[7]{Joint Institute of Nuclear Research, 141980~Dubna, Russia}
\address[8]{Institut f\"{u}r Kernphysik, Goethe-Universit\"{a}t, 60438~Frankfurt, Germany}
\address[9]{Excellence Cluster 'Origin and Structure of the Universe' , 85748~Garching, Germany}
\address[10]{Physik Department E12, Technische Universit\"{a}t M\"{u}nchen, 85748~Garching, Germany}
\address[11]{II.Physikalisches Institut, Justus Liebig Universit\"{a}t Giessen, 35392~Giessen, Germany}
\address[12]{Istituto Nazionale di Fisica Nucleare, Sezione di Milano, 20133~Milano, Italy}
\address[13]{Institute for Nuclear Research, Russian Academy of Science, 117312~Moscow, Russia}
\address[14]{Institute of Theoretical and Experimental Physics, 117218~Moscow, Russia}
\address[15]{Department of Physics, University of Cyprus, 1678~Nicosia, Cyprus}
\address[16]{Institut de Physique Nucl\'{e}aire (UMR 8608), CNRS/IN2P3 - Universit\'{e} Paris Sud, F-91406~Orsay Cedex, France}
\address[17]{Nuclear Physics Institute, Academy of Sciences of Czech Republic, 25068~Rez, Czech Republic}
\address[18]{LabCAF. F. F\'{\i}sica, Univ. de Santiago de Compostela, 15706~Santiago de Compostela, Spain} 
\address[a]{Also at Lawrence Berkeley National Laboratory, ~Berkeley, USA}
\address[b]{Also at ISEC Coimbra, ~Coimbra, Portugal}
\address[c]{Also at ExtreMe Matter Institute EMMI, 64291~Darmstadt, Germany}
\address[d]{Also at Technische Universit\"{a}t Dresden, 01062~Dresden, Germany}
\address[e]{Also at Dipartimento di Fisica, Universit\`{a} di Milano, 20133~Milano, Italy}
\address[f]{Also at Frederick University, 1036~Nicosia, Cyprus}
\address[g]{Also at Dipartimento di Fisica and INFN, Universit\`{a} di Torino, 10125~Torino, Italy}
\author[50]{\\
and A.V.~Sarantsev}
\address[50]{Petersburg Nuclear Physics Institute, Gatchina, Russia} 
%\vspace*{0.3cm}
%
%\vspace*{-3cm}
%
%
\cortext[mycorrespondingauthor]{\raggedright{}Corresponding authors: eliane.epple@ph.tum.de and laura.fabbietti@ph.tum.de}
\begin{abstract}
Employing the Bonn-Gatchina partial wave analysis framework (PWA), we have analyzed HADES data of the reaction $p(3.5GeV)+p\to$\pklo. 
This reaction might contain information about the kaonic cluster "$ppK^-$" via its decay into \plo.
Due to interference effects in our coherent description of the data, a hypothetical \KNN (or, specifically "$ppK^-$") cluster signal must not necessarily 
show up as a pronounced feature (e.g. a peak) 
in an invariant mass spectra like \plo. Our PWA analysis includes a variety of resonant and non-resonant intermediate states and delivers a good description of our data 
(various angular distributions and two-hadron invariant mass spectra) without a contribution of a \KNN cluster. 
At a confidence level of CL$_{s}$=95\% such a cluster can not contribute more than 2-12\% to the total cross section with a \pkl final state, 
which translates into a production cross-section between 0.7 $\mu b$ and 4.2 $\mu b$, respectively. 
The range of the upper limit depends on the assumed cluster mass, width and production process.
\end{abstract}
\begin{keyword}
\vspace*{-0.1cm}
%% keywords here, in the form: keyword \sep keyword       % PWA hadronic physics low  ernergy QCD strangeness
kaonic nuclei \sep antikaon-nucleon physics \sep $ppK^-$ \sep low energy QCD \sep partial wave analysis.
\PACS 11.80.Et \sep 13.75.Cs \sep 13.75.Jz \sep 21.30.Fe \sep  21.45.-v \sep 25.40.-h \sep 26.60.-c. 
\end{keyword}
%
%
%PACS NUMBERS:
%11.80.Et Partial-wave analysis
%13.75.Jz    Kaon-baryon interactions
%13.75.Cs    Nucleon- Nucleon interactions
%25.40.-h	Nucleon-induced reactions 
%21.30.Fe	Forces in hadronic systems and effective interactions
%21.45.-v	Few-body systems
%%24.10.-i Nuclear reaction models and methods
%26.60.-c	Nuclear matter aspects of neutron stars
%21.65.-f	Nuclear matter
%
%
\end{frontmatter}
%\begin{linenumbers}
%
%- - - - - - - - - - - - - - - - - - - - - - - - - - - - - - - - - - - - - - - - - - - - - - - - - - - - - - - - - - - -
\section{Introduction}
Quantum chromodynamics (QCD), in the low energy sector, is also a theory of hadrons. % \cite{.}. 
To describe such degrees of freedom one can use effective theories, which allow among other issues for the quantitative handling of meson baryon interactions. 
Restricting further the considerations to the SU(3) flavour sub-sector, the \KN interaction is of long standing interest \cite{Kaiser:1995eg,Oset:1997it,Lutz:2001yb}. 
Since the interaction of $\overline{K}$ (anti-kaon) and N (nucleons) was found to be attractive, particularly in the I=0 channel, speculation 
about the existence of bound systems, like the three-body state \KNNo, emerged \cite{Nogami1963288}. 
Most of the employed \KN potentials are based on the chiral meson-baryon (MB) interaction 
which is evaluated by coupled-channel calculations \cite{Kaiser:1995eg,Oset:1997it,Lutz:2001yb}. 
Experimentally, the \KN interaction in vacuum is probed by $K^{-}$p scattering experiments and by the investigation of kaonic atoms \cite{Martin:1980qe,Bazzi:2011zj}. 
The so-obtained potentials can be used as input to solve few-body problems like the \KNN bound state in Faddeev or variational calculations. 
An overview thereof is given in the tables of Refs.~\cite{Gal:2010eg,Gal:2013vx}. 
Nowadays, the discussion in this field concentrates mainly on the question whether the \KNN 
system is bound deeply (40--100 MeV \cite{Yamazaki:2002uh,Akaishi:2002bg,Shevchenko:2006xy,Ikeda:2007nz,Wycech:2008wf,Faber:2009jt}) 
or shallow (10--30 MeV \cite{Barnea:2012qa,Dote:2008hw,Ikeda:2010tk,Bayar:2012rk,Revai:2014twa}). 
The predicted widths generally exceed 40 MeV which complicates an experimental observation. 
Moreover, many theoretical works only discuss the mesonic ($YN\pi$) decay width of the \KNN state, while experimental analyses 
focus on the non-mesonic ($YN$) decay channel $p\Lambda$. 
\par
The discovery of kaonic nuclear bound states would deliver quantitative information 
about the strength of the binding of $\overline{K}$ to nucleons. 
This information could also help to understand a particular aspect of astronomical objects namely the interior of neutron stars \cite{Kaplan:1986yq,Nelson:1987dg}. 
In the inner core of these objects strange degrees of freedom could be favored to decrease 
the Fermi pressure by a condensation of kaons.
The discovered neutron stars with masses around 2M$_{\odot}$ \cite{Demorest:2010bx,Antoniadis:2013pzd} put, however, 
tight constraints on the stiffness of neutron star matter which is hardly compatible with a large fraction of condensed kaons \cite{Char:2014cja}, as this generally softens the equation of state. 
In this context, the depth of the \KN potential determines the maximum neutron star mass so that the study of kaonic nuclear bound states might help to answer the 
question on possible kaon condensation in neutron stars \cite{Kishimoto:1999yj}.
\par
So far, the discovery of a \KNN bound state was claimed by three experiments on the basis of measured
$p\Lambda$ invariant mass (M$_{p\Lambda}$) spectra.
The signal from FINUDA (M=$2255^{+6}_{-5}$(stat)$^{+3}_{-4}$(syst) \MeVo, $\Gamma$=$67^{+14}_{-11}$(stat)$^{+2}_{-3}$(syst) \MeVo) 
was reconstructed from stopped $K^{-}$ on thin nuclear targets \cite{Agnello:2005qj}.
The OBELIX signal (M=2212.2$\pm$4.9 \MeVo, $\Gamma<$24.4$\pm$8 \MeVo) was extracted from a multi-particle final state in $\overline{p}+^{4}$He 
reactions \cite{Bendiscioli:2007zza}, and the DISTO signal \cite{Yamazaki:2010mu} 
was obtained from the same reaction as in our work. 
This signal was selected by searching for deviations of the measured spectra from phase space distributions.  
A deviation in the M$_{p\Lambda}$ spectra was found and associated with a \KNN signal 
(M = 2267$\pm$3(stat)$\pm$5(syst) \MeVo, $\Gamma$ = 118$\pm$8(stat)$\pm$10(syst) \MeVo).  
This deviation was found only for the reaction p+p at a beam energy of 2.85 GeV, while absent at 2.5 GeV \cite{Kienle:2011mi}.
Since all the reported signals differ from each other and are, moreover, criticized \cite{Magas:2006fn,Ramos:2007zz,Mares:2006vk,Pandejee:2010ya,FinudaObel1}, 
an experimental confirmation of the theoretical predictions is far from being established.
Besides these findings, recently the LEPS and J-Parc E15 collaborations have reported on upper limits 
for the differential production cross section of a \KNN bound state in the $\gamma+d$ reaction and via the in-flight $^{3}$He($K^{−}, n$) reaction, respectively.
The reported upper limits depend on the assumed mass and width \cite{Tokiyasu:2013mwa,Hashimoto:2014cri}.  
\par
In the present work, open strangeness production via the reaction
\begin{flushleft} 
\vspace{-1.5cm}
\begin{flalign}\label{eq:pkl}
\begin{pspicture}(0,1)
  \psline[ArrowInside=-]{->}(3.85,-0.10)(3.85,-0.47)(4.5,-0.47)
\end{pspicture}
 p+p  \xrightarrow{3.5 GeV}p  + K^{+}+\Lambda & \\ 
                                        &\hspace{0.6cm} p+\pi^{-}                        \nonumber   
\end{flalign}
\end{flushleft}
is studied. This final state might reveal information of an intermediate production of the smallest kaonic nuclear cluster (\KNNo), named "$ppK^{-}$" and with 
the quantum numbers $J^{P}$=0$^{-}$, via its decay into a \pl pair. 
The underlying hypothesis for this reaction is that the possible formation of the \KNN cluster could proceed through the so-called $\Lambda(1405)$ 
doorway~\cite{Yamazaki:2002uh,Yamazaki:2007cs}. 
Under this hypothesis, the final state $\Lambda(1405)+p+K^{+}$ is formed in a first step while, subsequently, the final state interaction of the $\Lambda(1405)$ and the proton leads to 
the formation of a $\Lambda(1405)$-p bound state. This system is well known from variational calculations, where the density distributions of the \KNN constituents 
suggests that the meson-baryon structure of the $\Lambda(1405)$ is nearly unchanged in the three-body system, making it essentially a $\Lambda(1405)$-p bound state \citep{Dote:2008hw,Yamazaki:2007cs}. 
The groundwork of the analysis, presented here, was a measurement of the $\Lambda(1405)$ production cross section and its kinematics in the very same reaction 
\cite{Agakishiev:2012xk}. 
Conclusions out of this and the result, presented here, are discussed at the end of this work.
\par
Since several experiments have studied Reaction~(\ref{eq:pkl}) and discovered 
that it is dominated by the presence of N* resonances which decay into a $K^{+}\Lambda$ pair via 
$p+p \rightarrow p + (N^*  \rightarrow \Lambda  + K^+) $
\cite{Cleland:1984kb,AbdelBary:2010pc,AbdelSamad:2006qu,AbdElSamad:2010tz}, 
the dynamics of this process have to be modeled with care. 
A phase space model description of the data, without taking into account the dynamics of the process, is, thus, insufficient \cite{Epple:2012cq,Fabbietti:2013npl}. 
A very appropriate tool for such studies is a partial wave analysis, since it allows a description of the data taking into account intermediate resonant and non-resonant processes. 
In addition, it allows to include in the description the possible contribution of a kaonic nuclear cluster in a consistent way. 
One of the previous experiments has measured Process~(\ref{eq:pkl}) at 30 and 50 GeV/c incident momentum and performed a partial wave 
analysis (PWA) of the experimental data, finding significant structures in two amplitudes which could not unambiguously be assigned to specific quantum numbers \cite{Cleland:1984kb}. 
Beside this attempt, the work presented here constitutes the first application of a 
PWA to open strangeness production in p+p collisions in the few GeV region. 
In order to understand qualitatively how the different intermediate resonant and non-resonant processes contribute 
to the production of final state (\ref{eq:pkl}), we have utilized the Bonn-Gatchina PWA framework \cite{Anisovich:2006bc,Anisovich:2007zz}. 
This understanding is important, as these processes are the main contributions for the kaonic cluster search. 
\par
The analysis starts with the selection of those p+p collisions which produce the exclusive final state $pK^{+}\Lambda$.
Then a PWA with different intermediate N$^*$ resonant and non-resonant production processes is used to describe our data.
Any deviation of the so-obtained PWA-based model from the experimental data, particularly in the $p\Lambda$ mass spectrum, might indicate the
presence of a new signal. 
The observation of no significant deviation leads to the establishment of an upper limit on its production strength 
for a set of assumptions about the postulated "$ppK^{-}$" state.
%
%
%- - - - - - - - - - - - - - - - - - - - - - - - - - - - - - - - - - - - - - - - - - - - - - - - - - - - - - - - - - - -
\section{The Experiment}
The p+p experiment was carried out with the 
\textbf{h}igh-\textbf{a}cceptance \textbf{d}i-\textbf{e}lectron \textbf{s}pectrometer (\textbf{HADES}) 
at the SIS18 synchrotron (GSI Helmholtzzentrum in Darmstadt, Germany). 
Previous to this campaign, a Forward Wall hodoscope (FW) has been installed 7 m downstream the HADES target. 
It delivers a time information with a resolution of around 700 ps and covers polar angles from 0.33$^{\circ}$ to 7.17$^{\circ}$.
In the analyzed experiment, this detector was partially utilized to detect the decay proton from the $\Lambda$ in Reaction~(\ref{eq:pkl}). 
For more information about the experimental setup and particle identification we refer to Ref. \cite{Agakishiev:2009am}.
\par
In the present experiment, a beam of protons with 3.5 GeV kinetic energy and an intensity of $\approx$ 10$^{7}$ particles/s was 
incident on a liquid hydrogen target with a density of
0.35 g/cm$^2$ corresponding to a total interaction probability of about 0.7\%.
The total recorded statistic contains 1.2$\times$10$^{9}$ events 
which fulfill the first-level trigger condition demanding three hits in the TOF detectors. 
\par
Out of these events the final state of Reaction~(\ref{eq:pkl}) has been selected.  
Two data-sets have been defined for the exclusive analysis: one, where all four particles were detected by the main HADES
spectrometer (called HADES data-set) and one, where the secondary proton from the $\Lambda$ decay hit the FW, while the other 
three particles were detected by HADES (called WALL data-set). 
In both cases a kinematic fit was applied to select the \pkl final state exclusively, and the kaon mass distribution was used to reject part of the remaining background \cite{Fabbietti:2013npl}.
The main source of physical background after the event selection comes from the reaction 
\begin{equation}\label{eq:pks}
p+p \rightarrow p+ K^+ + \Sigma ^0,
\end{equation}
that contributes to the selected events with 1\% and 3\% in the HADES and WALL data-sets respectively. 
Additional background originates from the mis-identification of pions and protons as kaons. This background amounts to 6.5\% (HADES case), and 11.7\% (WALL case). 
After the event selection a total number of 22,000 \pkl events\footnote{13,000 events from the HADES data-set and 9,000 events from the WALL data-set.} remains for the analysis which is a sufficiently large statistic.
%
%- - - - - - - - - - - - - - - - - - - - - - - - - - - - - - - - - - - - - - - - - - - - - - - - - - - - - - - - - - - -
\section{The Partial Wave Analysis}
The analysis of the measured $pK^{+}\Lambda$ events was performed with the Bonn-Gatchina partial wave analysis framework \cite{Anisovich:2006bc,Anisovich:2007zz}.
This PWA allows to decompose the baryon-baryon scattering amplitude into separate sub-processes characterized by different intermediate states. 
For the investigated process, where three particles with four-momenta $q_i$ are produced from a collision of two particles with 
four-momenta $k_i$, the production cross-section can be written as \cite{Ermakov:2011at} 
\begin{flalign}
&d\sigma=\frac{(2\pi)^4 |A|^2}{4 |\bf{k}| \sqrt{s}} d\Phi_3(P,q_1,q_2,q_3),\\
&\textrm{with }P=k_1 +k_2.
\end{flalign}
Here, $A$ is the transition amplitude, $| \bf{k}| $ the beam momentum in the p-p center-of-mass system, $\sqrt{s}$
the center-of-mass energy of the reaction and $d\Phi_3$ the phase space element of the three-particle final state. 
The transition amplitude $A$ is decomposed into partial waves according to \cite{Ermakov:2011at} 
\begin{flalign}\label{eq:Atot}
A=&\sum_{\alpha} A^{\alpha}_{tr}Q^{in}_{\mu_1...\mu_J}(S,L,J)A_{2b}^{\alpha}(S_2,L_2,J_2)(s_i) \nonumber \\ 
     &\times Q^{fin}_{\mu _1...\mu _J}(i,S_2,L_2,J_2,S',L',J).
\end{flalign}
Where $S,\,L,\,J$ represent the combined spin, orbital momentum and total angular momentum of the initial p+p system.
%It is defined by the spectroscopic notation $^{2S+1} L_J$, 
For our experiment, we only 
consider states with $J<3$ which translates in the following allowed initial states: $^{2S+1} L_J$=$\lbrace ^1S_0,\, ^3P_0,\, ^3P_1,\, ^3P_2,\, ^1D_2,\, ^3F_2 \rbrace$.
\par
$A^{\alpha}_{tr}$ is the transition amplitude from the initial to the intermediate quasi-two-body state, where the index $\alpha$
runs over all allowed combinations of the final state quantum numbers. 
As our data were taken at a fixed energy the amplitude is parametrized as follows \cite{ASarants_PrivComm} 
\begin{equation}\label{eq:transAmpl}
A^{\alpha} _{tr}= a^{\alpha}_1  e^{ia_2^{\alpha}}.
\end{equation}
Here, the complex amplitude is determined by a strength $a^{\alpha}_1$ and a phase $a_2^{\alpha}$. 
\par
The production of the \pkl final state might proceed either directly or via intermediate N* resonances.
In the former case a $p\Lambda$ subsystem is constructed, and the kaon is treated with respect to this system. In the latter case, the
$K^{+}$ and the $\Lambda$ form the N* resonance, and the proton is treated with respect to this system. 
$s_i$ is the invariant mass of the two-particle subsystem: $s_i=(P-q_i)^2$, given $q_i$ the four-momentum of the third particle. 
In our case only the two particle systems \pl and $K^{+}\Lambda$ are considered. 
The quantum numbers $S_2,\,L_2,\,J_2$ contain the information about the subsystem, while the 
third particle $K^{+}$ or proton is assigned with the quantum numbers $S',\,L',\,J$, respectively. 
The quantities $Q^{in}_{\mu_1...\mu_J}(S,L,J)$ and $Q^{fin}_{\mu_1...\mu_J}(i,S_2,L_2,J_2,S',L',J)$ are the spin-momentum operators of the initial 
and final states respectively, which amongst others contain the angular dependence of the scattering amplitude \cite{Anisovich:2006bc,Anisovich:2007zz,Anisovich:2004zz}. 
\par
The amplitude $A_{2b}^{\alpha}(S_2,L_2,J_2)(s_i)$ of the two-body subsystem in Equation~(\ref{eq:Atot}) contains either: 
\\
the elastic scattering of the proton and the $\Lambda$ in non resonant production processes 
with the parametrization \cite{Ermakov:2011at,ASarants_PrivComm}
\begin{equation}
A^{\beta}_{2b}(s_{p \Lambda})=\frac{r^{\beta}_{p\Lambda}a^{\beta}_{p \Lambda} \sqrt{s_{p \Lambda}}}
{1-\frac{1}{2}r^{\beta}_{p\Lambda}q^2 a^{\beta}_{p\Lambda} + 
\frac{iq a^{\beta}_{p \Lambda} q^{2L_2}}{F(q,r^{\beta}_{p\Lambda},L_2)}},
\end{equation}
with q being the relative momentum between the p and the $\Lambda$, $a^{\beta}_{p\Lambda}$ the p$\Lambda$ 
scattering length, $r^{\beta}$ the effective range of the p$\Lambda$ system, 
and $F(q,r,L_2)$ the Blatt-Weisskopf form factor \cite{Anisovich:2004zz}; 
the index $\beta$ represents a subset of the index $\alpha$ and accounts only for the possible combinations of quantum numbers describing the two-body sub-system $S_2,\,L_2,\,J_2$;\\
or $A_{2b}$ contains the production of N* resonances parametrized by a relativistic
Breit-Wigner amplitude \cite{Anisovich:2006bc} 
\begin{equation}\label{eq:BW}
A^{\beta}_{2b}(s_{K^{+} \Lambda})= \frac{g_{K^{+}\Lambda}}{M^2-s_{K^{+} \Lambda}-iM\Gamma_{Tot}},
\end{equation}
with $M$ and $\Gamma_{Tot}$ being the pole mass and the width of the resonance, respectively. The factor $g_{K^{+}\Lambda}$ is the decay coupling of the resonance to the 
$K^{+}\Lambda$ system. 
\par
The Bonn-Gatchina PWA performs a global fit of the data which implies that external resonance parameters are needed. In fact, the parameters 
$a^{\alpha}_1$ and $a^{\alpha}_2$ in Eq.~\ref{eq:transAmpl} are the only free fit parameters. 
The parameters of resonances with an observed decay into the $K^{+}\Lambda$ channel and masses accessible in the probed energy regime are taken
from Ref.~\cite{Beringer:1900zz}. 
These are the following states: N(1650)$\frac{1}{2}^{-}$, N(1710)$\frac{1}{2}^{+}$, N(1720)$\frac{3}{2}^{+}$, N(1875)$\frac{3}{2}^{-}$, N(1880)$\frac{1}{2}^{+}$, N(1895)$\frac{1}{2}^{-}$, and N(1900)$\frac{3}{2}^{+}$. Of which the N(1880) and N(1895) only have a two star rating in the PDG. 
The input waves build an ansatz for the PWA which is fitted on an event-by-event basis to the data. 
The angular dependencies of the partial wave amplitudes are constructed using the four-vectors measured inside of detector acceptance. 
The fitted parameters $a_{1}^{\alpha}$ and $a_{2}^{\alpha}$ in Eq.~(\ref{eq:transAmpl}) are optimized to gain the maximum of the
likelihood function. 
This value is calculated as the product of probabilities for all measured events normalized to the total cross section obtained within the HADES acceptance. 
The retrieved solutions allow us to reconstruct the multi dimensional detector acceptance using a set of full-scale phase space simulations.
\par
\begin{figure}[t!]
\begin{center}
\includegraphics[width=\textwidth]{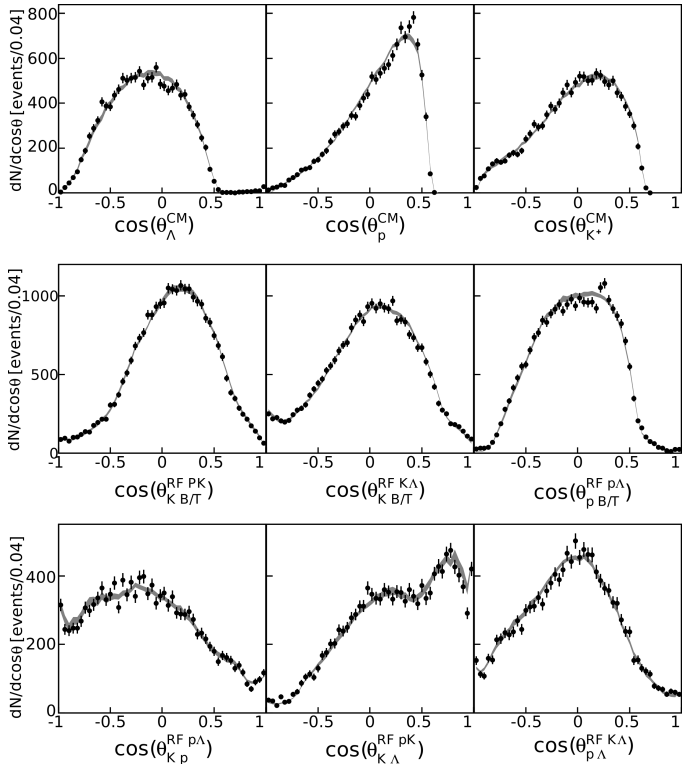}
\caption[]{Angular correlations for the \pkl final state, within the detector acceptance, shown for the HADES data-set. Black dots are the experimental data with their statistical uncertainty while the gray band shows the four best solutions of the PWA and displays their systematic differences. The upper index at the angle indicates the rest frame (RF) in which the angle is displayed. The lower index names the two particles between which the angle is evaluated. CM stands for the center-of-mass system. B and T denote the beam and target vectors, respectively. The observables are: CMS angles (upper row), Gottfied-Jackson angles (middle), and helicity angles (lower row). For further details on the observables see Ref.~\cite{Agakishiev:2011qw}.
}
\label{Fig:AngleH}
\end{center}
\end{figure}
\begin{figure}[hbt]
\begin{center}
\includegraphics[width=\textwidth]{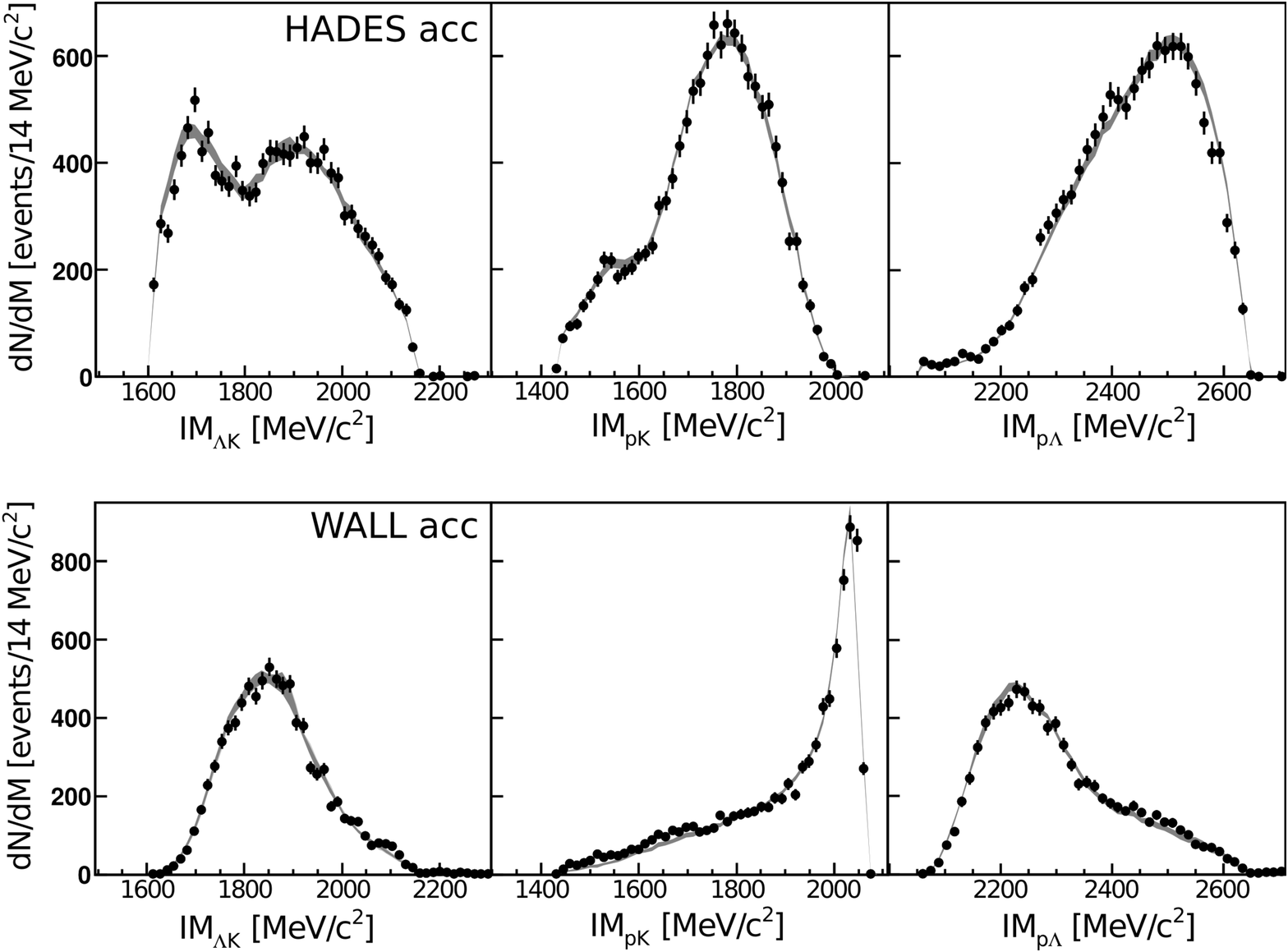}
\caption[]{Two-particle mass distributions for the \pkl final state, within the detector acceptance, shown for the HADES (upper panels) and WALL data-sets (lower panels), respectively. Black dots are the experimental data with their statistical uncertainty while the gray band shows the four best solutions of the PWA and displays their systematic differences.}
\label{Fig:Mass}
\end{center}
\end{figure}

To account for the large uncertainties on the existence and properties of part of the listed resonances, different ansatzes have been fitted to the data. 
Table~\ref{table:NonResSyst} contains ten versions of non-resonant production waves (left part) and twelve versions of N* resonances (right part) which were used as 
intermediate states. Their combination yields 120 different ansatzes that were fitted to the data. 
The goodness of a fit is characterized by the negative of the log-likelihood value that has been minimized in the fitting procedure.
%\begin{widetext}
%-------------------------------------------------------
\begin{table}[bht]
%\begin{table*}[t]
\begin{center}
\caption{Different sets of non-resonant and resonant waves used as PWA input. The non resonant waves are described by an $(p\Lambda)$ isobar with the quantum numbers 
written in the spectroscopic notation $^{(2S+1)}L_J$ and displayed in the brackets. %which contains the spin S, orbital angular momentum L,and total angular momentum J
Additionally, the kaon can have various angular momenta with respect to the $p\Lambda$ system in each displayed wave.}
\label{table:NonResSyst} 
\begin{tabular}{rlrl}  
 \toprule
No. &  Non-resonant  & No. &  Resonant   \\
      &  contributions   &       &   contributions \\
 \midrule   
    0    & no waves                        &     0 &   N(1650), N(1710), N(1720)\\
    1    &  $(^{1}S_{0})$             &     1 &   No. 0+N(1900)\\
    2    & No. 1$ +(^{3}S_{1})$  &     2 &   No. 0+N(1895)\\
    3    & No. 2$ +(^{1}P_{1})$  &     3 &   No. 0+N(1880)\\
    4    & No. 3$ +(^{3}P_{0})$  &     4 &   No. 0+N(1875)\\
    5    & No. 4$ +(^{3}P_{1})$  &     5 &   No. 0+N(1900), N(1880)\\
    6    & No. 5$ +(^{3}P_{2})$  &     6 &   No. 0+N(1900), N(1895)\\
    7    & No. 6$ +(^{1}D_{2})$  &     7 &   No. 0+N(1900), N(1875)\\
    8    & No. 7$ +(^{3}D_{1})$  &     8 &   No. 0+N(1895), N(1880)\\
    9    & No. 8$ +(^{3}D_{2})$  &     9 &   No. 0+N(1895), N(1875)\\
       &                                      &   10 &   No. 0+N(1880), N(1875)\\ 
       &                                      &    11 &   all resonances w/o No. 0\\
 \bottomrule
\end{tabular}
\end{center}
%\end{table*}
\end{table}
%-------------------------------------------------------
%\end{widetext}
To account for the systematic uncertainty on the choice of the included waves in the fit result, the four best solutions of this systematic variation 
were taken as the result of the fit.
These solutions are: No. 8/1, 8/3, 9/6, and 8/8 (Non-resonant/Resonant combination), of which solution 9/6 had the best log-likelihood value. 
The fact that these combinations describe the data equally well, although the resonances used in the ansatz of the PWA were different, shows that 
the two data-sets are not sufficient for the PWA to determine the unique resonance contributions to the considered final state. 
To exhibit the quality of the four PWA solutions, the theoretical differential cross sections, calculated within the HADES acceptance, are scaled 
to the experimental data in Figures~\ref{Fig:AngleH} and~\ref{Fig:Mass}, 
which show several angle and mass distributions. The gray band includes the four best solutions and displays their systematic differences 
which are small despite their content differs quite strongly from one another. 
The agreement between data and the PWA solutions is excellent. 
To test effects that might bias the result of the PWA fit, several checks have been performed. These are discussed in Refs.~\cite{EppleMESON,Epple:2014}. 
One check shows that the fraction of background events in  

\noindent
the data does not decrease the predictive power of the
fit \cite{EppleMESON} and the other check was performed to test whether an 
unknown signal that is in the data might bias the result of the PWA \cite{Epple:2014}. 
%
%- - - - - - - - - - - - - - - - - - - - - - - - - - - - - - - - - - - - - - - - - - - - - - - - - - - - - - - - - - - -
\section{The Hypothesis Tests and the Upper Limit}
\begin{figure}[htb]
\floatbox[{\capbeside\thisfloatsetup{capbesideposition={right,top},capbesidewidth=.36\textwidth}}]{figure}[0.56\textwidth]
{\caption{The local $p_{0}$ value and the equivalent significance for different masses of $p\Lambda$. It is calculated based on the mass spectra from the HADES and WALL data. The gray hatched range is due to the systematic uncertainty between the four best solutions of the PWA.}\label{Fig:LocalP0}}
{\includegraphics[width=.6\textwidth]{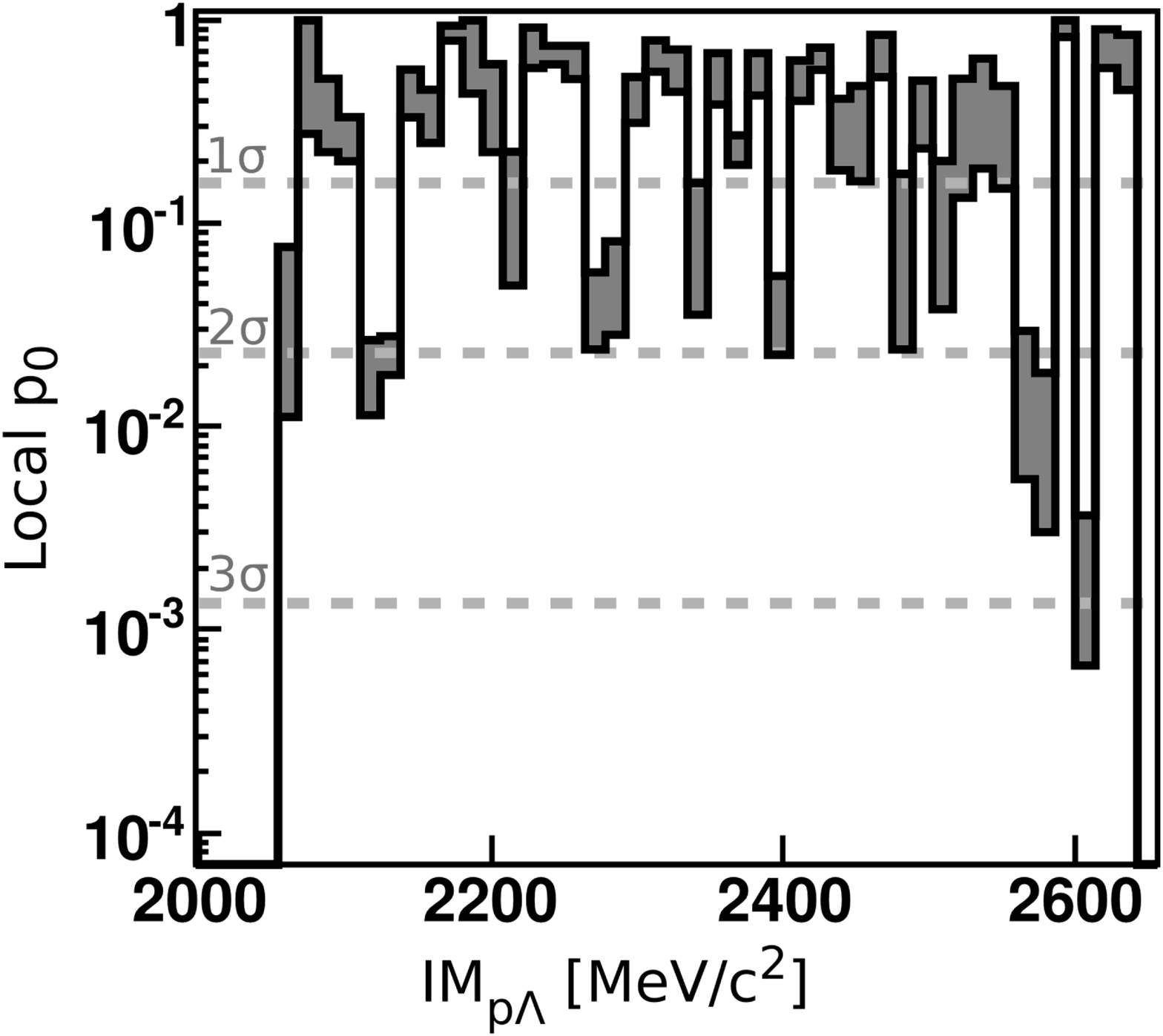}}
\end{figure}

The four best PWA solutions were used as a null hypothesis $H_{0}$ for the existence of the kaonic nuclear bound state with its decay into \plo. 
A significant deviation of the data from the PWA results might indicate the presence of an additional signal, like the \KNNo.
The discrepancy between the measured data and the null hypothesis as a function of the $p\Lambda$ 
invariant mass was determined based on a local $p_{0}$-value \cite{Epple:2014}. The combined result of this hypothesis test 
including both mass spectra (HADES and WALL data) is shown in Figure~\ref{Fig:LocalP0}. 
The different $p_{0}$-values of the four PWA solutions were combined to a gray band. 
%- - - - - - - - - - - - - - - - - - - - - - - - - - - - - - - - - - - - - - - - -
%- - - - - - - - - - - - - - - - - - - - - - - - - - - - - - - - - - - - - - - - -
The local $p_{0}$-value and its according equivalent significance, shown in units of $n\sigma$, shows a good agreement between $H_{0}$ and the 
data.\footnote{A correct hypothesis will produce p-values uniformly distributed between 0 and 1. If the H$_{0}$ hypothesis is false the p-values should be distributed more likely at very small values. This is a necessary condition for the presence of a new signal in the data.}
In the possible mass range of the kaonic nuclear bound state 2054-2370 \MeV the agreement is always within $2\sigma$. 
Hence, the data are consistent with $H_{0}$ and we do not observe any significant contribution of a yet unknown signal, like the \KNNo, to the data. 
This conclusion does also hold for the separate local p$_{0}$-values for the HADES and WALL data, as shown in Ref.~\cite{Epple:2014}.
\par
\begin{figure*}[t]
\begin{center}
\includegraphics[width=0.8\textwidth]{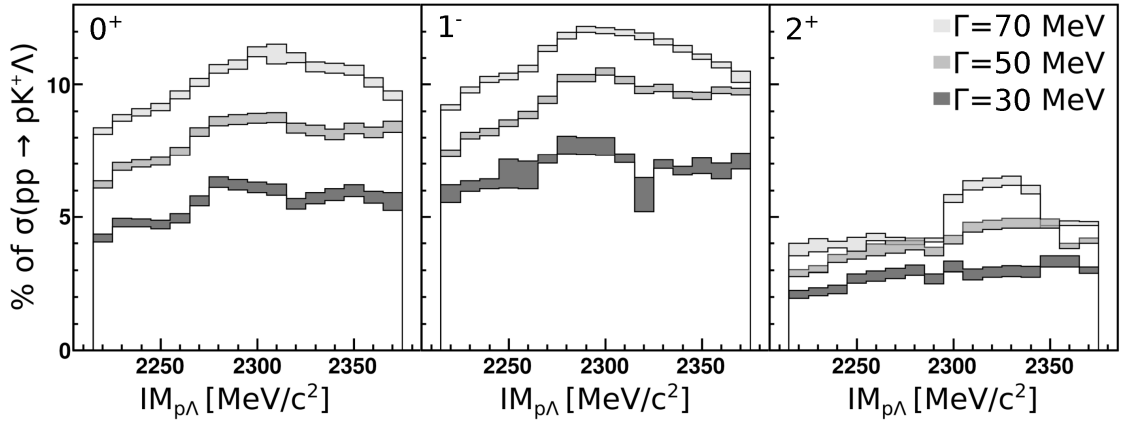}
\caption[]{The upper limit of the \KNN production strength at a $CL_{s}$ value of 95\%. The limit is quoted in \% of the total \pkl production cross section of the investigated reaction. 
The three figures display the limits for the three transition amplitudes in which the cluster can be produced (0$^+$,1$^-$ and 2$^+$).}
\label{Fig:UppLimit}
\end{center}
\end{figure*}
In a next step the data were tested against several signal hypotheses to determine an upper limit of the \KNN contribution to the data.
For that purpose, the \KNN signal has been included as a wave to the PWA solution. 
The \KNN was parametrized as a Breit-Wigner in the \pl system according to Eq. \ref{eq:BW}. 
As the mass and width of the state are not known, we have tested signals with masses of 2220-2370 \MeV in steps of 10 \MeVo.
For the width, values of 30, 50, and 70 \MeV were combined with each mass. 
%To test the presence of the bound state, 
The \KNN state with the quantum numbers $J^{P}$=0$^{-}$ \cite{Nogami1963288,Ikeda:2007nz} can be produced out of three initial p+p configurations: 
$^{2S+1}L_{J}$=$\lbrace ^{1}S_{0}$, $^{3}P_{1}$, $^{1}D_{2}\rbrace$ which corresponds to waves with $J^{P}$=0$^{+}$, 1$^{-}$, and $2^{+}$, respectively.
The \KNN has been included in the fit in these three waves separately. In the new PWA solution the amplitude $a_{1}^{\alpha}$ in Eq.~(\ref{eq:transAmpl}) was increased step-wise, 
while the phase of the \KNN wave was freely varied.
This phase determines the interference patterns that are caused by the wave. 
Due to this effect a larger signal can be included into the solution with a less pronounced appearance in the mass spectrum.
\par
The upper limit was determined with the CL$_{s}$ method (confidence level of the signal), which is ideal for setting signal limits in case of low sensitivity 
\cite{Beringer:1900zz,Junk:1999kv,Read:2000ru,Read:2002hq}, and was calculated based on the \pl invariant mass distribution.
An amplitude strength that corresponds approximately to the less than 5\% most likely outcomes of the measured data, given the signal hypothesis, was rejected by the test 
(CL$_{s}$-value higher than 95\%). 
This amplitude scan was repeated for each of the four PWA solutions and the highest of the four limits is presented in Figure~\ref{Fig:UppLimit}. 
It shows an upper limit of the \KNN cluster production as a function of the hypothetical mass in \% of the total \pkl production cross section. 
This cross section was determined to $\sigma_{pK^{+}\Lambda}(3.5GeV) = 38.12\pm0.43^{+3.55}_{-2.83}\pm2.67{-}2.86$ $\mu$b 
(statistical, systematical and normalization uncertainty are given with the result, as well as the contribution from background that needs to be subtracted) \cite{EppleMESON} 
and allows thus to quote the 
upper limit of a \KNN bound state cross section, which reads 1.8-3.9$\mu$b, 2.1-4.2$\mu$b, and 0.7-2.1$\mu$b, respectively.
%
%- - - - - - - - - - - - - - - - - - - - - - - - - - - - - - - - - - - - - - - - - - - - - - - - - - - - - - - - - - - -
\section{Summary and Conclusion}
We have performed a partial wave analysis (PWA) of \pkl events to search for signals of the hypothetical kaonic nuclear cluster "$ppK^-$". 
The two analyzed data-sets do not allow to pin down the exact contribution of the N* resonances to the \pkl final state. 
Our approach, together with a more comprehensive analysis of many \pkl data-sets at several beam energies, could, however, be the right way to resolve this issue.  
The description of the data by PWA solutions, including only known sources, is satisfactory, so that no convincing argument requesting a new signal is needed. 
Adding, nevertheless, an assumed \KNN signal into the PWA we tested quantitatively a signal hypothesis against the data.
This test was performed at a CL$_s$ level of 95\%. Due to this limit we have accepted the about 5\% most unlikely data outcomes, given the model,  
to set the upper limit. The limit on the kaonic cluster production strength in the mass range M=2220-2370 \MeV and assuming widths of $\Gamma$=30, 50, and 70 \MeV is given for the three possible production waves $J^{P}$=$\lbrace0^{+}, 1^{-}, 2^{+}\rbrace$. 
The limits lie between 5-11\% (0$^{+}$), 6-12\% (1$^{-}$), and 2-6\% (2$^{+}$) of the total \pkl production cross section. 
Using the extracted cross section from $\sigma_{pK^{+}\Lambda}(3.5GeV)$, this translates into upper limits of 1.8-3.9~$\mu$b, 2.1-4.2~$\mu$b, and 
0.7-2.1~$\mu$b for the \KNN cluster production cross section, respectively.
These limits are not comparable to searches \cite{Tokiyasu:2013mwa,Hashimoto:2014cri} which rely on incoherent analyses, as in these analyses 
a cross section is defined as an observed, rather than a produced yield. 
We emphasize, therefore, that our PWA analysis includes, for the first time, interference between the waves. 
This allows to include a larger fraction of produced \KNN cluster without a visible appearance e.g. as peaks in the \pl mass spectrum. 
We also note that our upper limit is given specifically for the \pl decay channel of the kaonic nuclear cluster with the quantum numbers $J^{P}$=0$^{-}$. 
\par
The upper limit of about~4~$\mu$b can be compared to the extracted production cross section of the $\Lambda(1405)$ of 
about~10~$\mu$b from the same experiment \cite{Agakishiev:2012xk}. 
This connects, also for the first time, two quantities that constrain the predicted dominance of the 
$\Lambda(1405)$ doorway scenario for the kaonic cluster formation in p+p reactions \cite{Yamazaki:2007cs}. 
Our results put at question scenarios where the probability of the $\Lambda$(1405)-p final state to form a \KNN cluster is very large. 
\par
With this work there are, meanwhile, as many reports of upper limits as signals published. 
This leaves us at a situation where the experimentalists rather create new puzzles than solve the theoretical controversy. 
Thus, in order to test low energy QCD and determine the strength of the \KN interaction, more data and more advanced analysis techniques like the introduced PWA are certainly needed.
%
%- - - - - - - - - - - - - - - - - - - - - - - - - - - - - - - - - - - - - - - - - - - - - - - - - - - - - - - - - - - -
\section*{Acknowledgments}
The authors kindly thank F. Beaujean for the discussion on statistical analysis.
The HADES collaboration gratefully acknowledges the support by the grants: 
PTDC/FIS/113339/2009 LIP Coimbra, NCN Poland, 2013/10/M/ST2/00042 Helmholtz 
Alliance HA216/EMMI GSI Darmstadt, VH-NG-823, Helmholtz Alliance HA216/EMMI 
TU Darmstadt, 283286, 05P12CRGHE HZDR Dresden, Helmholtz Alliance HA216/EMMI, 
HIC for FAIR (LOEWE), GSI F\&E Goethe-University, Frankfurt VH-NG-330, BMBF 06MT7180 
TU München, Garching BMBF:05P12RGGHM JLU Giessen, Giessen UCY/3411-23100, 
University Cyprus CNRS/IN2P3, IPN Orsay, Orsay MSMT LG 12007, AS CR M100481202, 
GACR 13-06759S NPI AS CR, Rez EU Contract No. HP3-283286
%
%2013/10/M/ST2/00042 Poland
%- - - - - - - - - - - - - - - - - - - - - - - - - - - - - - - - - - - - - - - - - - - - - - - - - - - - - - - - - - - -
\bibliography{KNN_lib2}
\end{document}